\documentstyle[psbox]{WORKSHOP}
\begin{document}

\title{
RXTE Observations of New Black Hole Candidates \\
XTE J1752--223 and MAXI J1659--152\\ 
{\large\sf  -- Timing and Spectral Evolution and BH Masses --} 
}

\author{
Nikolai Shaposhnikov,$^{1,2,3}$ Jean H. Swank,$^3$ Craig Markwardt,$^3$ \\ 
and  Hans Krimm$^{2,4}$
\\[12pt]  
%
$^1$  CRESST/University of Maryland, Department of Astronomy, College Park MD, 20742\\
$^2$  Goddard Space Flight Center, NASA, Astrophysics Science Division, Greenbelt MD 20771\\
$^3$   CRESST/Universities Space Research Association, Columbia MD, 21044\\
%
{\it E-mail(NS): nikolai.v.shaposhnikov@nasa.gov} 
}

\abst{
We report the most recent results of our analysis of X-ray monitoring of new Galactic 
Black Hole (BH) candidates  XTE J1752--225 and MAXI J1659--152 performed by
Rossi X-ray Timing Explorer ({\it RXTE}). We investigate various aspects of the
RXTE data including energy and power spectra, variability  energy distribution
and phase lags between soft and hard energy bands. The sources generally exhibit the spectral
states and evolution expected from an accreting stellar mass BH. The energy distribution of different
variability  components show that the aperiodic noise has a spectrum  consistently softer with respect to the total
rms spectrum, while the spectrum of the quasi-periodic (QPO) features is harder. Particularly interesting behavior is
observed in phase lags. Namely, XTE J1753--223 shows
that QPO  in the hard band lags the QPO in the soft band. This is opposite to what was previously reported 
in other bright  BH candidates and also  found in our analysis from MAXI J1659--152. We
report the results of BH mass estimations using the spectral-timing correlation scaling technique.
Namely, we obtain the BH masses of 9.5$\pm$1.5 and 20$\pm$3 solar masses for XTE J1752--223 and
MAXI J1659--152 correspondingly.
}

\kword{accretion---black hole physics---stars: individual --- XTE J1752--223, MAXI J1659--152}

\maketitle
\thispagestyle{empty}

\section{Introduction}

During 2009 and 2010  the {\it Rossi X-ray Timing Explorer (RXTE)} performed extensive
monitoring campaigns on the newly discovered galactic black hole candidates
 XTE J1752--223 (Markwardt et al. 2009, J1752 hereafter)  and MAXI J1659--152 (Negoro et al. 2010,
 J1659 hereafter).  General evolution of the discovery outburst of J1752 is reported 
 in Shaposhnikov et al. (2010). Here we present further analysis of J1752, including power spectra,
variability energy distribution (rms spectra) and time lags.
 Synergy of these {\it RXTE} data products provides deep insight into 
unexplained phenomena related to accreting BHs
 such as aperiodic and quasi-periodic variability and 
 non-thermal emission.  
We also present the evolution of J1659 throughout 
the discovery outburst. 

The main feature of PDS during state transition is a low frequency  quasi-periodic oscillation (QPO)
with centroid  frequency evolving from below 0.1 Hz up to 10 Hz as a source moves from LHS to HSS.
Usually, QPOs are observed on top of a broad-band variability which has a broken power
law shape roughly constant at low frequencies (also known as flat-top noise or white red noise).
The nature of these variability components is still in debate. 
It is essential to study all observational aspect of X-ray data to solve the puzzle of
variability in BH sources. 

We investigate the correlation of the spectral 
index with the QPO frequency .
 The correlation pattern shows clear saturation of the index 
as the source moves along the transition from the hard to soft state. In the framework
of the Bulk Motion Comptonization (BMC) scenario this effect is interpreted
as a signature of the converging inflow onto a BH horizon (see Shaposhnikov \& 
Titarchuk, 2009, ST09 hereafter, and references therein). 
J1659 provides further evidence for ubiquity of the index saturation effect in  BH candidates.
In Shaposhnikov et al. (2010) 
we showed that the evolution of the rms spectrum during
the state transition in J1752 can be explained by changing partial contributions
of the thermal Comptonization and BMC in upscattering soft photons.
Titarchuk \& Shaposhnikov (2010) also explained the behavior of the high energy cutoff observed in XTE J1550-564
 within the BMC framework.

Correlations of the QPO frequency with spectral parameters provide
a means to measure BH masses and to estimate source distances (ST09).
In this Paper we summarize the results of the scaling method for J1752 and
J1659. Our results indicate that the BH mass in J1659 may be
the highest among the Galactic BH sources measured so far, i.e.  about 20 solar masses.
Given a possible binary period of 2.4 hours (Kuulkers et al. 2010), this
system may present an exotic high BH mass ultra-compact BH binary. 

\begin{figure}[t]
\centering
\psbox[xsize=8cm]{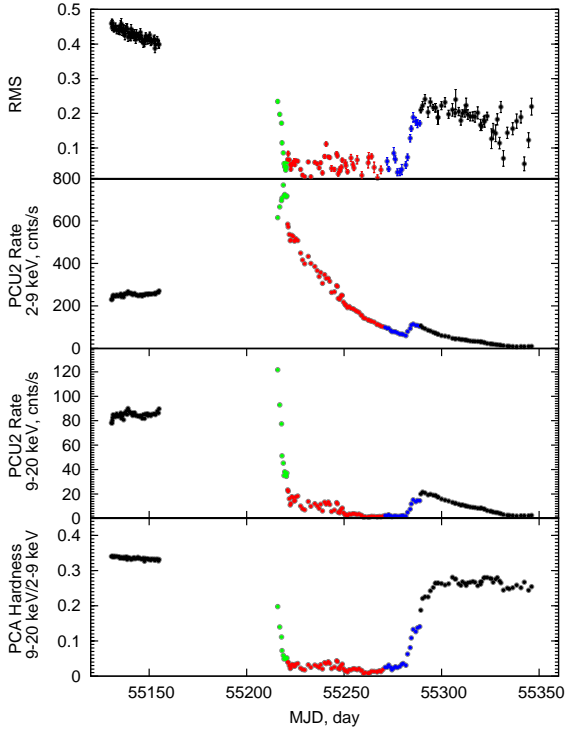}
\caption{ Evolution of the J1752 outburst. {From the top to the bottom:} Total rms variability (0.01-64.0 Hz) in the 
whole PCA range (i.e. no channel selection made), PCA lightcurves in 2-9 keV and 9-20 keV energy ranges and 
the spectral hardness versus time.}
\label{evol1752}
\end{figure}

\begin{figure}[t]
\centering
\psbox[xsize=8cm]{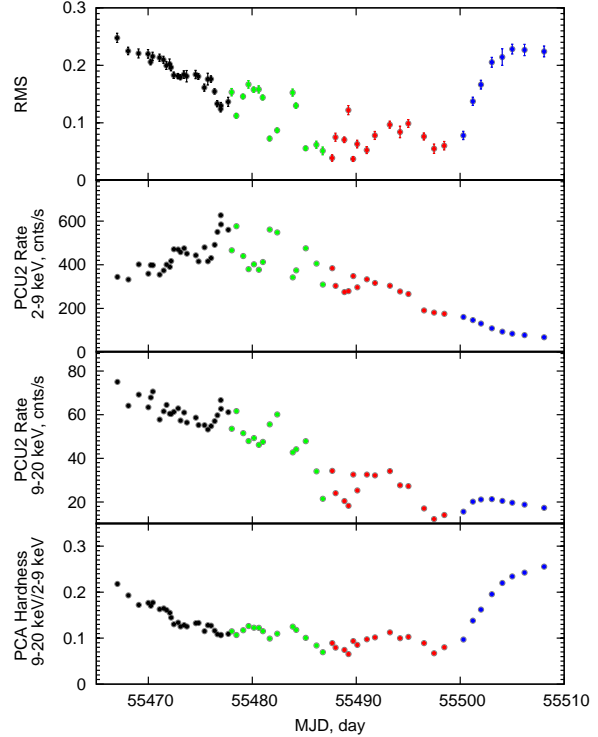}
\caption{ The same as Figure \ref{evol1752} for the J1659 outburst evolution.}
\end{figure}

In the next section we briefly describe 
the discovery and {\it RXTE} observations of J1752 and J1659. In  Section 3 we
discuss general spectral evolution of the outbursts.
In the Section 4 we present a detailed analysis of the J1752
using the Fourier techniques, i.e. power spectra, 
rms spectra and time lags. In Section 5 we examine the correlations between 
the QPO and the spectral parameters in J1659 and summarize the BH mass measurements using
the correlation scaling method. Conclusions follow in Section 6.

\section{Sources, Observations and Data Analysis}
\subsection{Discovery of J1752 and J1659}
J1752 was discovered during the {\it RXTE}/PCA Galactic Bulge Scans performed 
on October 23, 2009 17:52 UT. 
The outburst lasted for about 8 months and exhibited a uniquely
long rise low-hard state (LHS), transition to the high-soft state (HSS) and a reverse transition to the decay LHS 
(see Shaposhnikov et al. (2010) for details). 

{\it Swift}/BAT triggered on an unidentified source on September 25, 2010 08:05:05 UT which
was preliminary identified as Gamma-ray burst (Mangano et al. 2010). {\it MAXI} detection
and observations provided correct identification as a Galactic X-ray transient (Negoro et al. 2010).
{\it RXTE} pointed observations started on September 28, 2010 00:43 UT and
revealed phenomenology consistent with the source being a new BH candidate (Kalamkar et al. 2010a).

\subsection{Observations}

{\it RXTE} have covered both outbursts with intensive daily monitoring
programs. During the first half of the outburst from J1659, a very frequent,
 2-3 per day observation strategy was utilized revealing
rich timing phenomenology. Unfortunately, the {\it RXTE} 
observations during both campaigns were not continuous 
 due to the Sun observational
constrains. Specifically, a significant part of the LHS and IS in J1752 were not
 observed beginning from MJD 55155 until MJD 55215.
For the same reason,  the latest part of the J1659 outburst
starting from MJD 55508 was not covered by {\it RXTE}
pointing observations. However, most of the soft-to-hard transition
was observed, allowing study of the QPO evolution and subsequent
BH mass measurement (see Section 5).

\subsection{Data analysis}

First we perform a standard spectral analysis by extracting
deadtime and background normalized PCA spectra using
the Standard2 data mode, providing 16 sec time resolution data
in 129 energy channels. For the spectral analysis we utilized
the data from PCU2 only. We fit the spectra in the 3-45 keV energy range
in XSPEC (Arnaud 1996) with a Generic Comptonization model 
(XSPEC BMC model) modified by interstellar absorption and, 
in most cases of the LHS and IS observations, by a high 
energy cutoff component. Also, a gaussian at $\sim$ 6.4 keV with a width of 0.5-1.5 keV was added
to account for possible contribution of the iron K$_{\alpha}$ emission line.

Next, to study the variability of the source emission
 we calculate the Fourier Power Density Spectra (PDS) for all
observation using the high time resolution data modes in the whole PCA
range. The PDSs were then integrated to get the total root-mean-square 
(rms) variability and fitted using a sum of Lorentzians to find QPO frequencies. 

We also investigated the energy dependence of variability in different frequency 
ranges using the method of Fourier Resolved spectroscopy (Revnivtsev et al. 1999).
Phase lags are calculated as an argument of a cross-spectrum
between lightcurves in 1.5-5 keV and 10-12 keV energy bands . For presentation, the phase lags shown
throughout the manuscript are normalized to $2\pi$, The time lags are obtained 
by dividing the phase lags by the Fourier frequency. A positive lag means that the harder 
variability lags that at softer energies. 

\section{Evolution}
\label{evol}

We present the evolution of J1752 and J1659 throughout the outbursts
in Figures 1 and 2 respectively. The top panels show evolution of the fractional rms variability, 
source fluxes in energy ranges of 2-9 keV and 9-20 keV, inferred from
our spectral model, and their ratio, commonly referred to as a hardness ratio. 
Data for the LHS and the HSS are plotted in black and red, while
rise and decay IS are shown in green and blue correspondingly.

There are number of differences in the outburst behavior between J1752 and J1659.
First, as it was already mentioned, the J1659
outburst was about 5 times shorter than the event form J1752. J1659 have not
shown a long hard state. In fact, after 3 days after the discovery the source was already in
the hard IS, showing a 1.6 Hz QPO (the PDS for this observation is shown in the upper left panel of Fig. 6).  
The state transition episodes in J1752 are dynamically much shorter than
the LHS and HSS stages and also faster than IS episodes observed in J1659.
In fact, the IS episodes in J1659 seem to be a part of a gradual source evolution.
Furthermore, the lowest hardness value achieved by J1659 (0.06$\sim$0.07, see the bottom
panel in Fig. 2) is 10 times higher than that shown by J1752. In fact, the power law component during the second half
of the HSS in J1752 became so weak that the spectrum could be considered completely
thermal. In opposite, in J1659 the power law remained relatively strong throughout the
entire event. 

\begin{figure*}[t]
\centering
\hspace{-0.1in}\psbox[xsize=4.35cm,rotate=r]{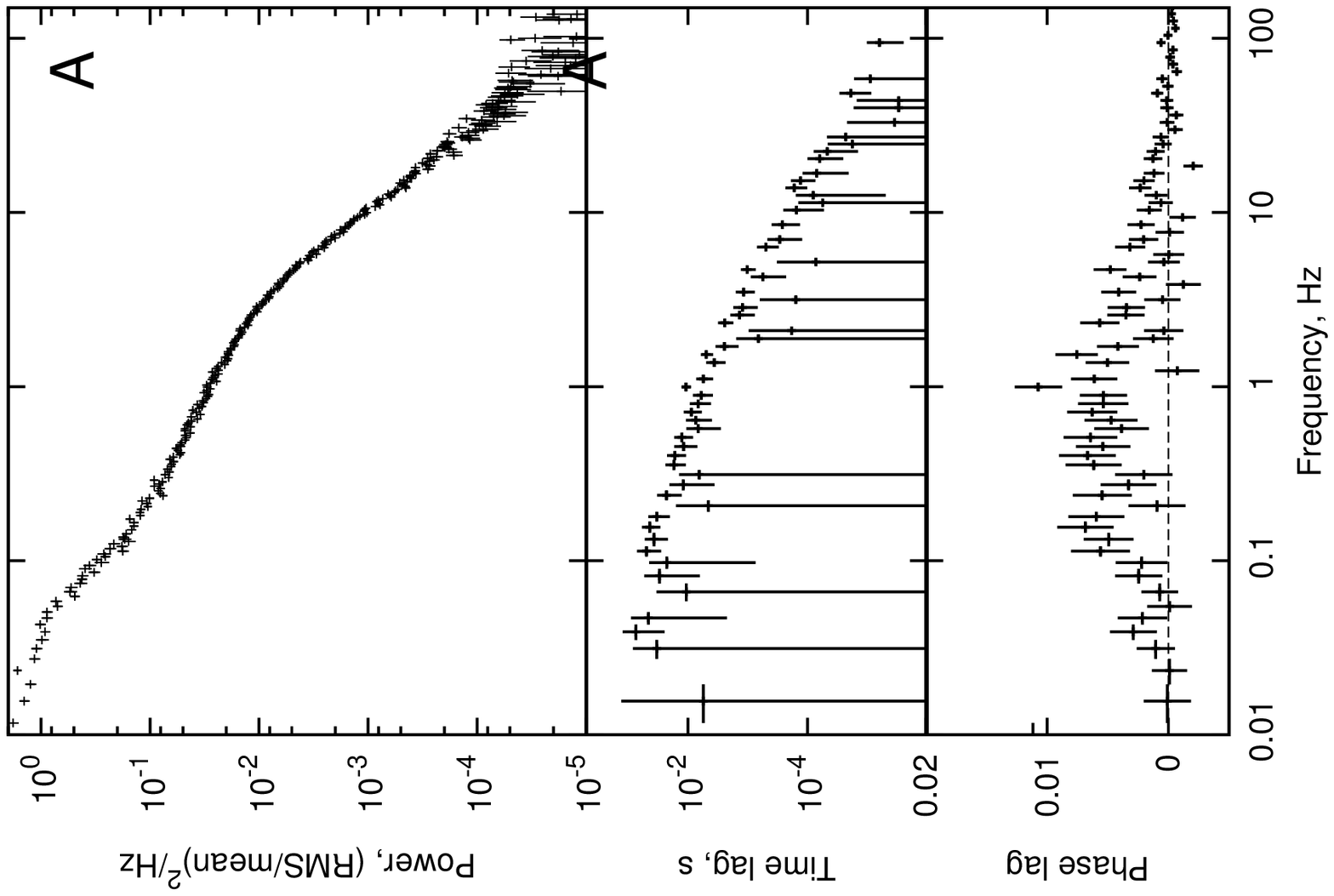}
\hspace{-0.1in}\psbox[xsize=4.35cm,rotate=r]{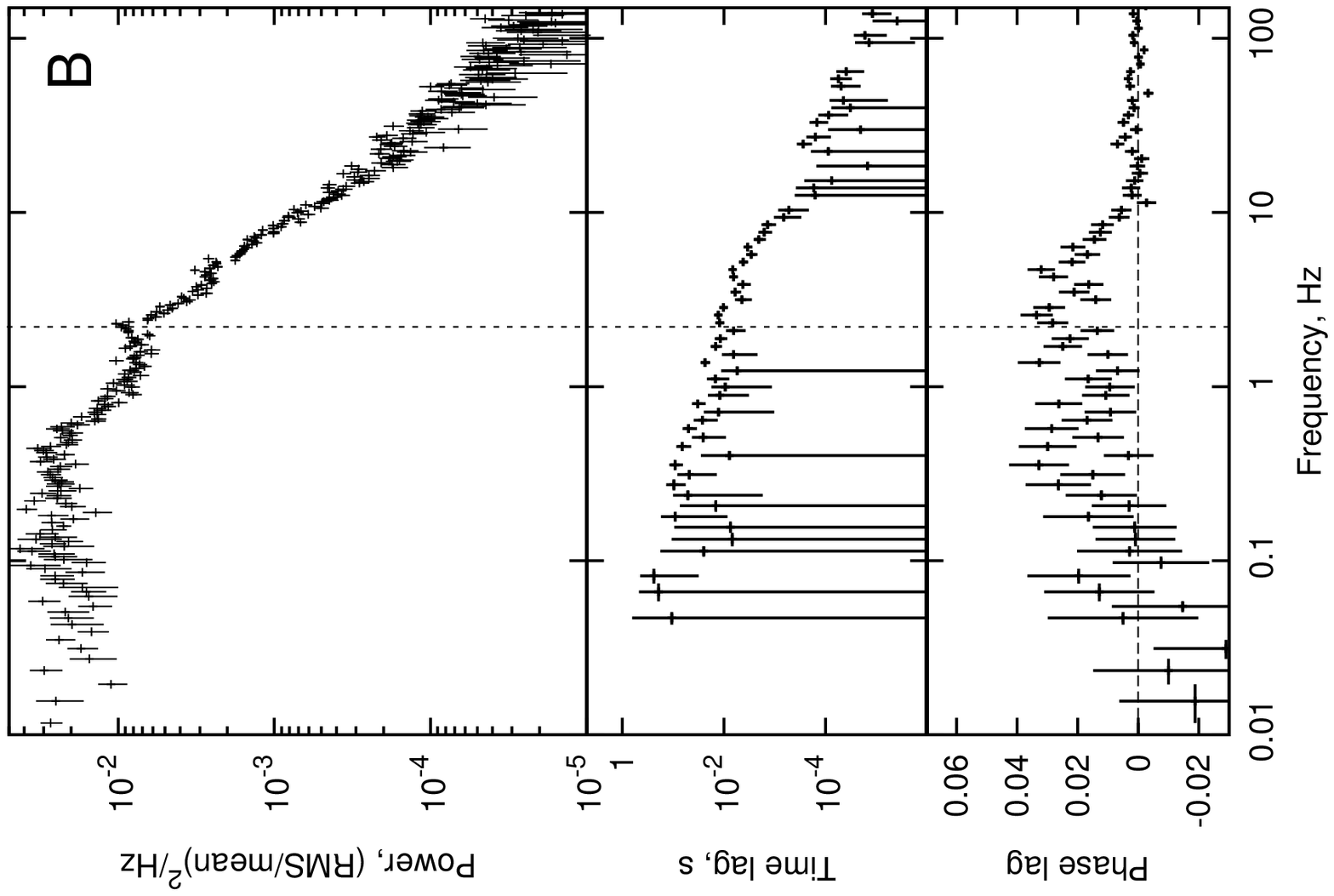}
\hspace{-0.1in}\psbox[xsize=4.35cm,rotate=r]{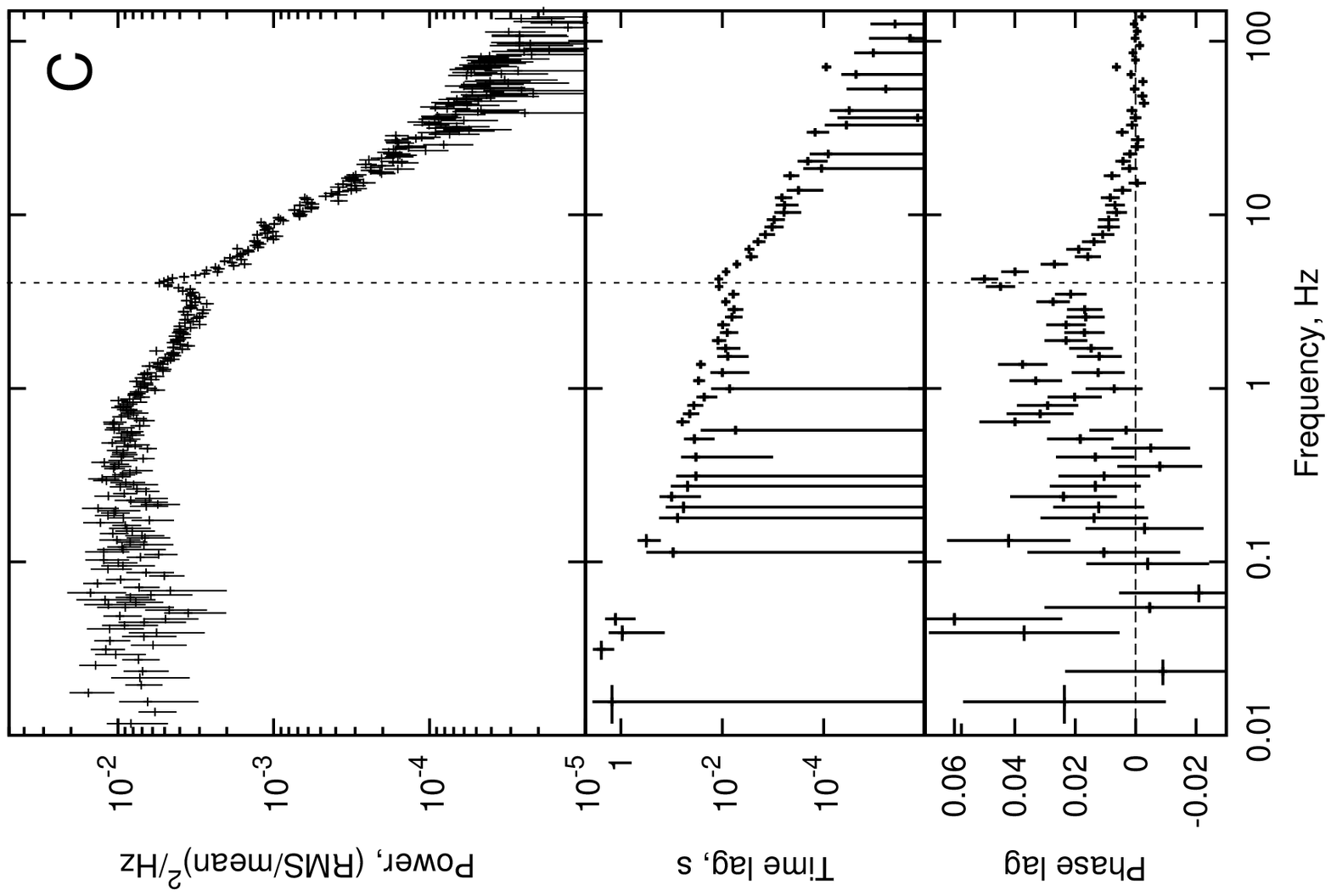}
\hspace{-0.1in}\psbox[xsize=4.35cm,rotate=r]{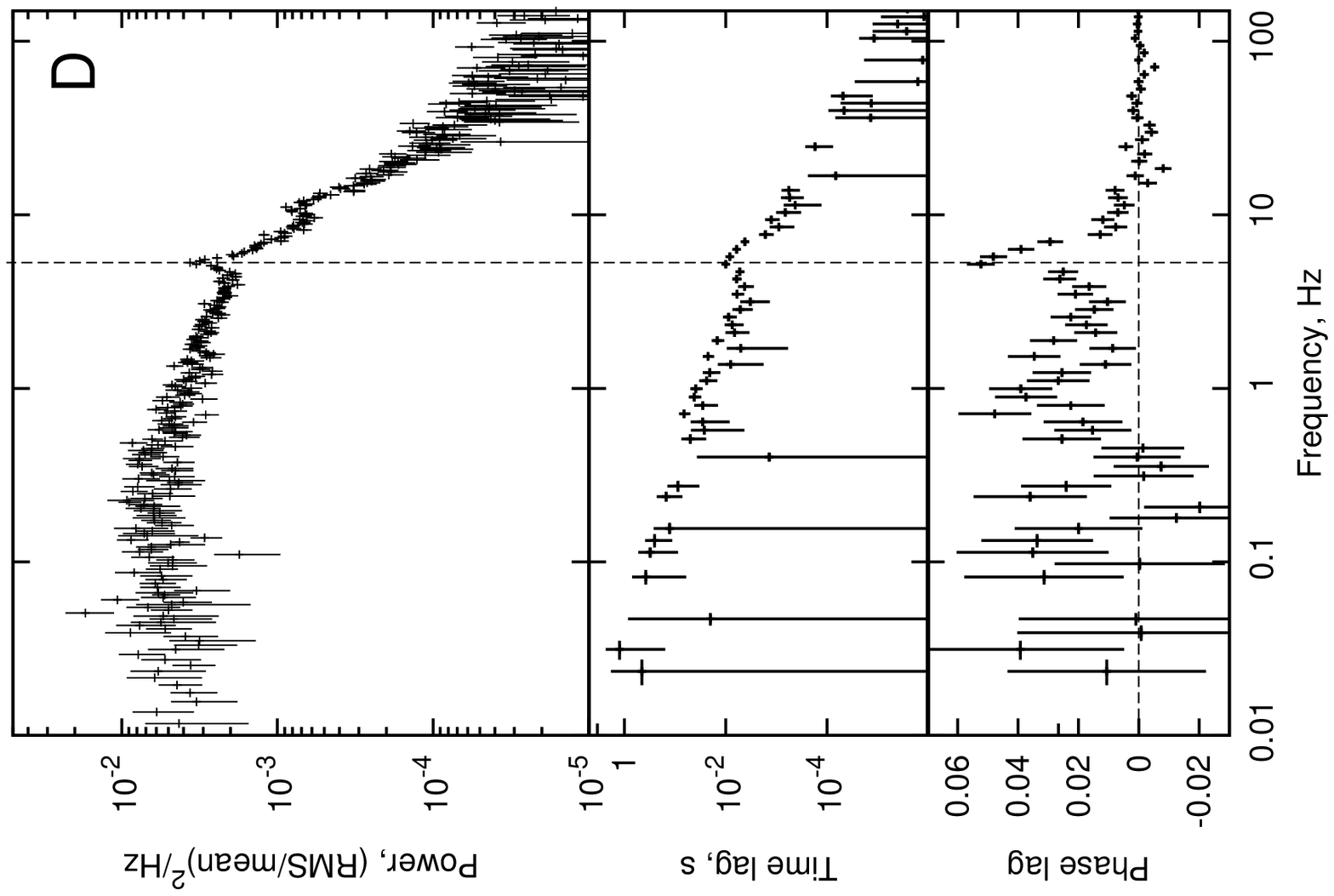}

\psbox[xsize=4.1cm,rotate=r]{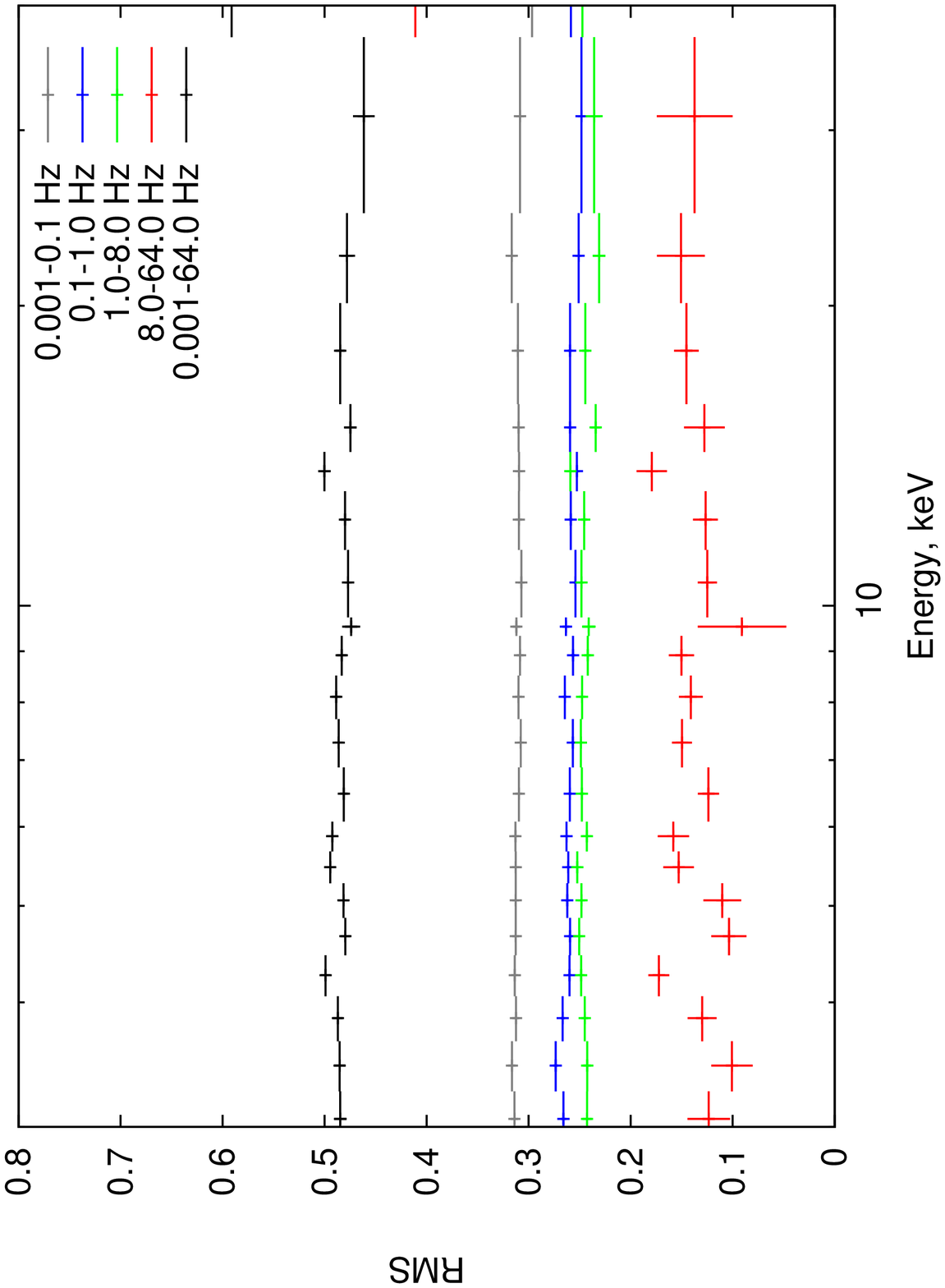}
\psbox[xsize=4.1cm,rotate=r]{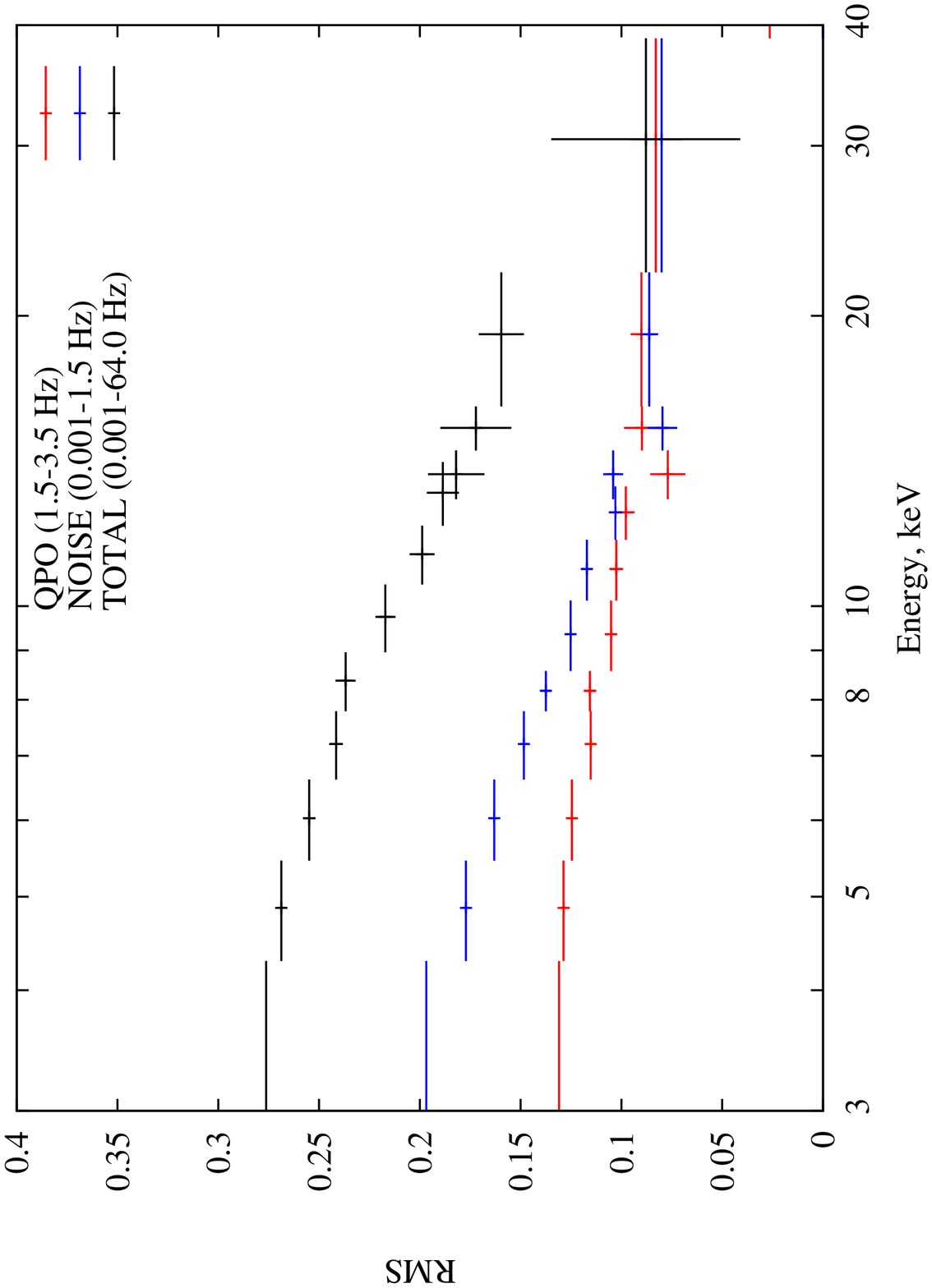}
\psbox[xsize=4.1cm,rotate=r]{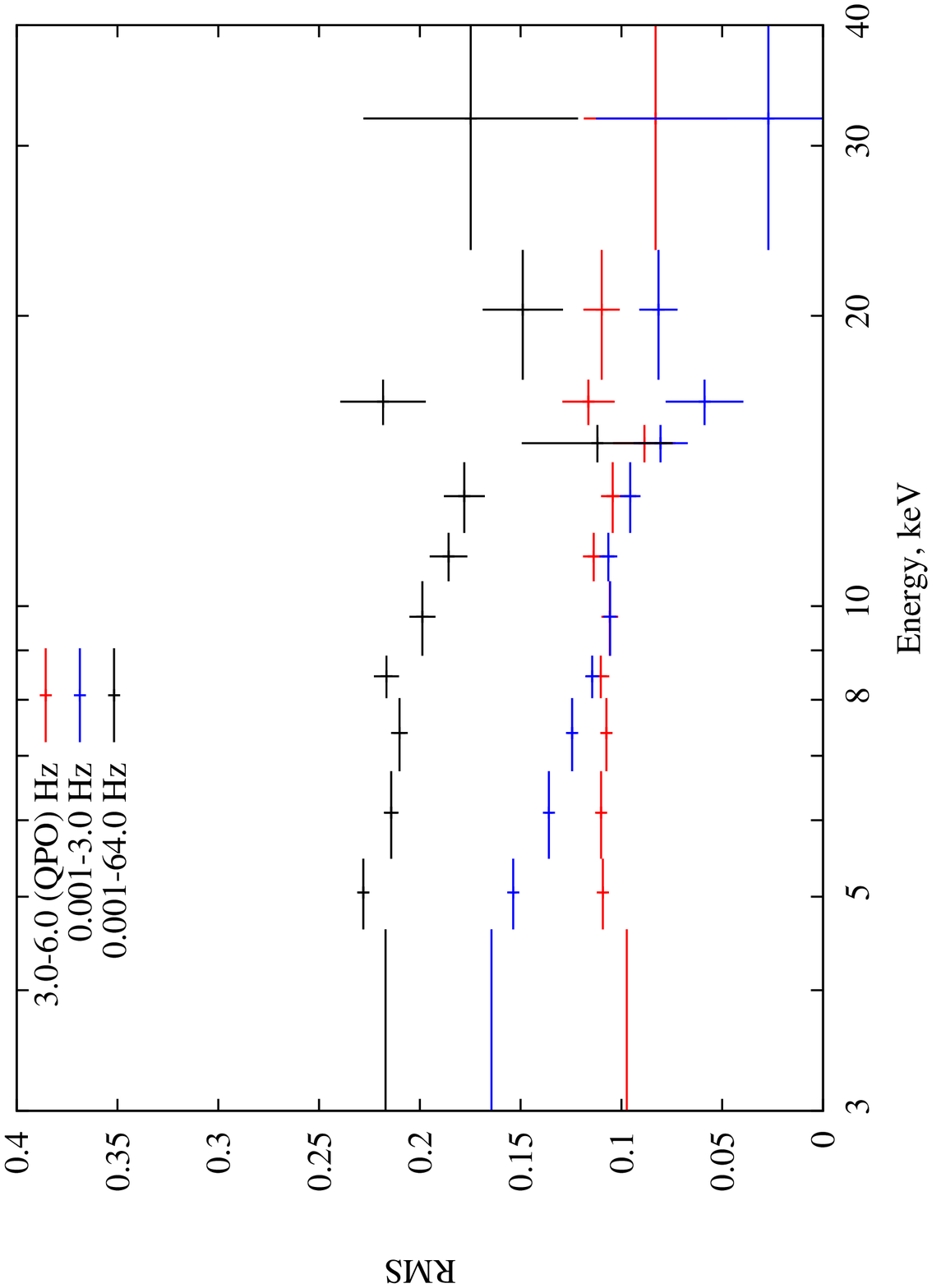}
\psbox[xsize=4.1cm,rotate=r]{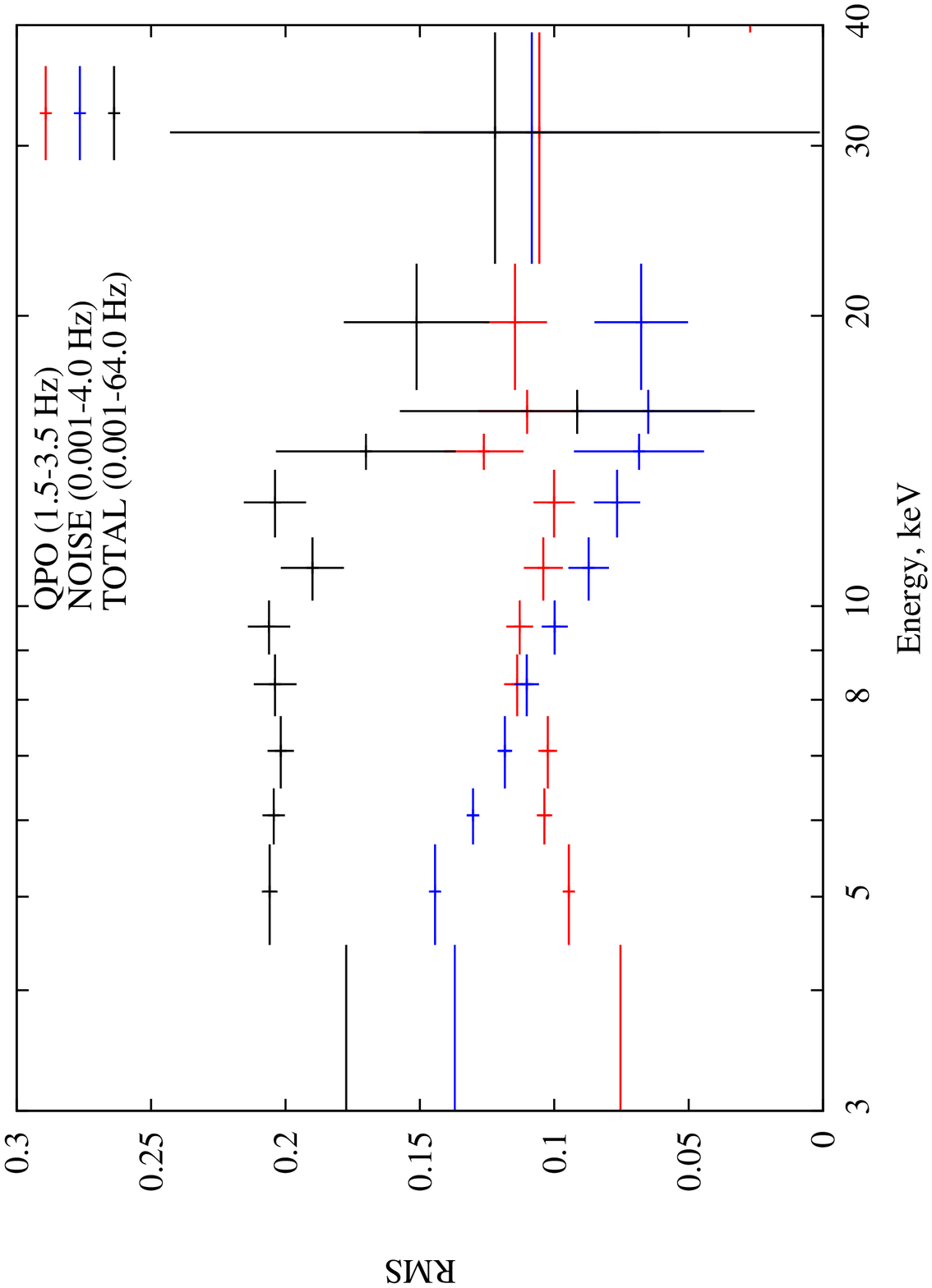}
\caption{  Fourier analysis of  four {\it RXTE} data sets during  the LHS and the IS in J1752.
From left to right diagrams present analysis of the following {\it RXTE} observations: 94044-07-01-00 (combined with -000 and -010 data subsets, start time 2009-10-26 14:52 UT) , 94331-01-06-00 (2010-01-19 21:36 UT), 94331-01-06-01 (2010-01-20 22:48 UT) and 94331-01-06-02 (2010-01-21 20:52 UT). The top diagrams show the power spectra, time and phase lags from  top to bottom. The phase lags are calculated between the lightcurves corresponding to 1.5-5 keV and 10-12 keV energy ranges. The lower figures show the energy dependence of rms variability in frequency bands related to the aperiodic and QPO variability. }
\end{figure*}

\section{PDS, RMS Spectra and Time Lags}

In this Section we focus on the variability properties shown by J1752 and J1659 during an initial hard-to-soft
state transitions. We present a detailed analysis of different power spectral components including
rms spectra and time lags. 

In Figures 3, 4 and 5  we show three different type of {\it RXTE} data products based on
Fourier analysis: PDS, rms spectra and times lags. Figures 3 and 4 show the results 
for six observations of J1752 covering the range of spectral states from the extreme LHS (panels A in Fig. 3) 
through the hard IS showing Type C QPOs (panels B,C and D) to the soft IS showing type B and A 
QPOs (Fig. 4). The first observation therefore belongs to the start of LHS (black points in Fig. 1),
while the rest of the observations belong to the IS (green data points in Fig. 1).

The following important observations can be made in regard to the evolution
shown on the Figures 3 and 4. As the total variability drops from $\sim$ 50 \%
in the LHS to below 10 \% at the end of IS (see Shaposhnikov et al. 2010) the rms spectra
evolve from constant to decreasing with energy at the start of the IS and to the increasing
with energy during the  soft IS (see Fig. 4). Rms spectra of the QPO are consistently harder than
both the the total rms spectra and the rms spectra of the broad-band noise.
\label{timing}

\begin{figure}[t]
\centering
\hspace{-0.1in}\psbox[xsize=4.2cm,rotate=r]{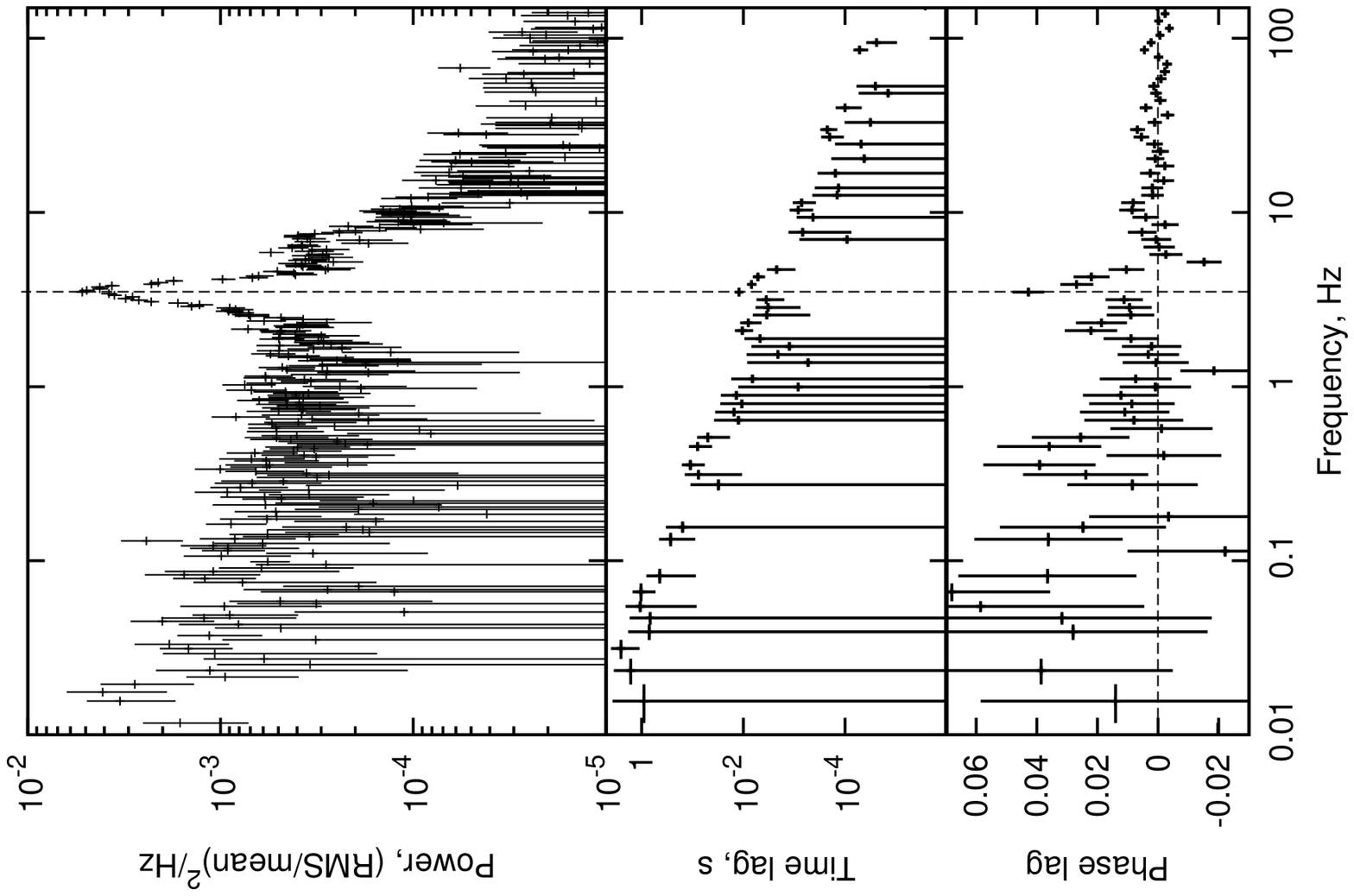}
\hspace{-0.1in}\psbox[xsize=4.2cm,rotate=r]{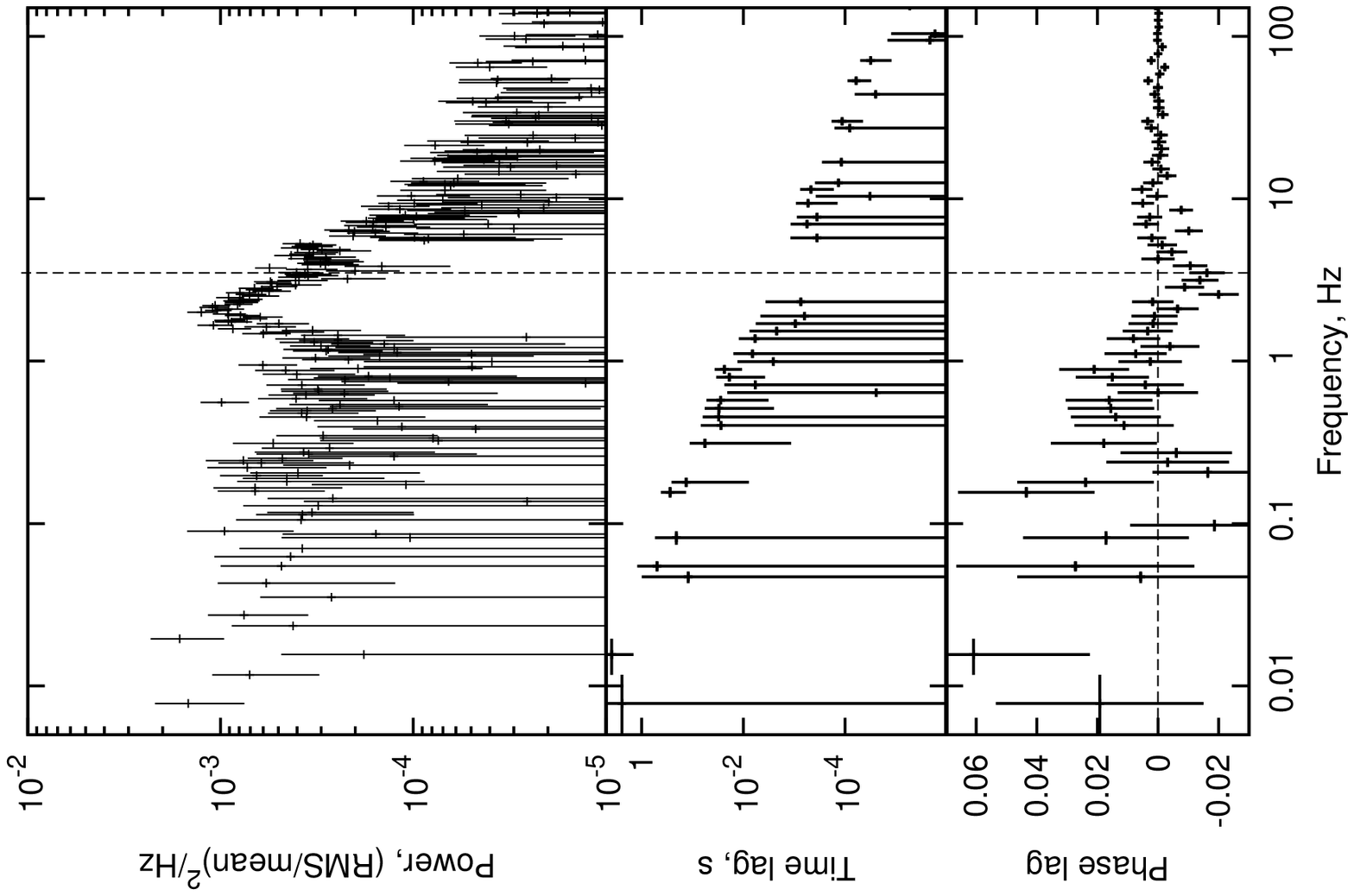}

\psbox[xsize=4.0cm,rotate=r]{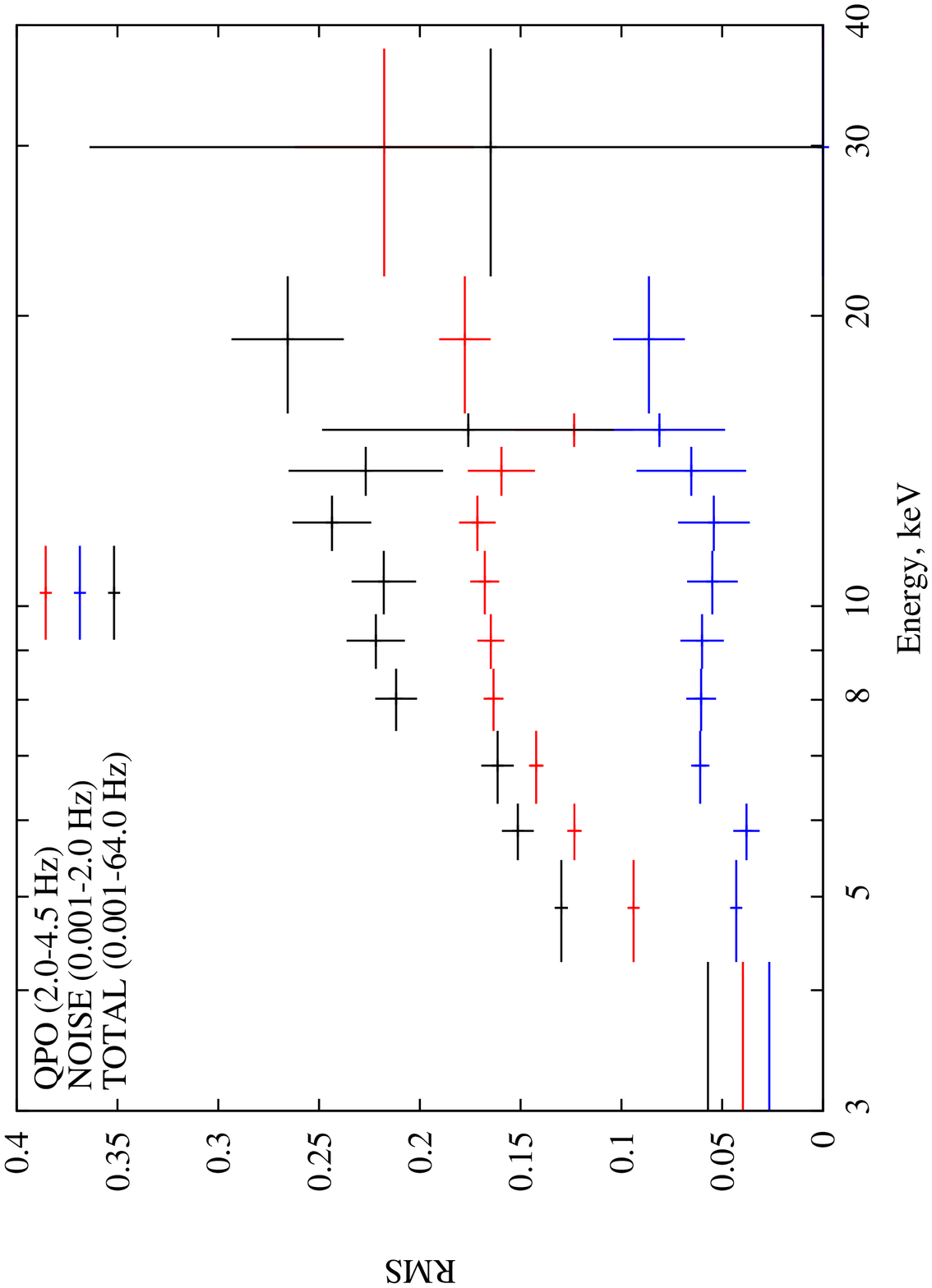}
\psbox[xsize=4.0cm,rotate=r]{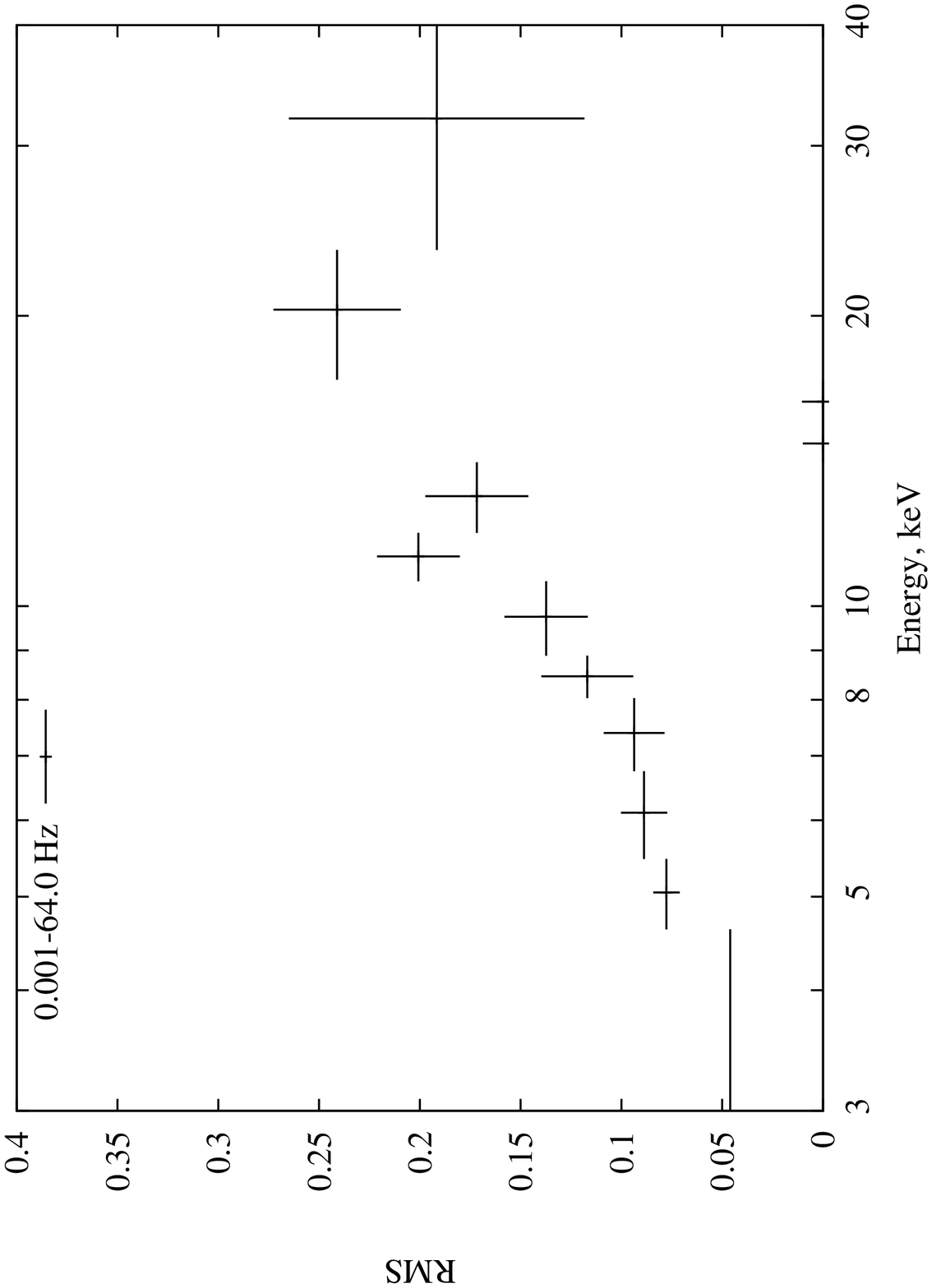}
\caption{Same as Figure 3 for J1752 observations 95360-01-01-00 (left panels, start time  2010-01-22 18:57 UT) and 95360-01-01-02 (right panels, 2010-01-24 16:04 UT) 
which were performed the soft IS episode. }
\end{figure}

\begin{figure}[t]
\centering
\psbox[xsize=8cm,rotate=r]{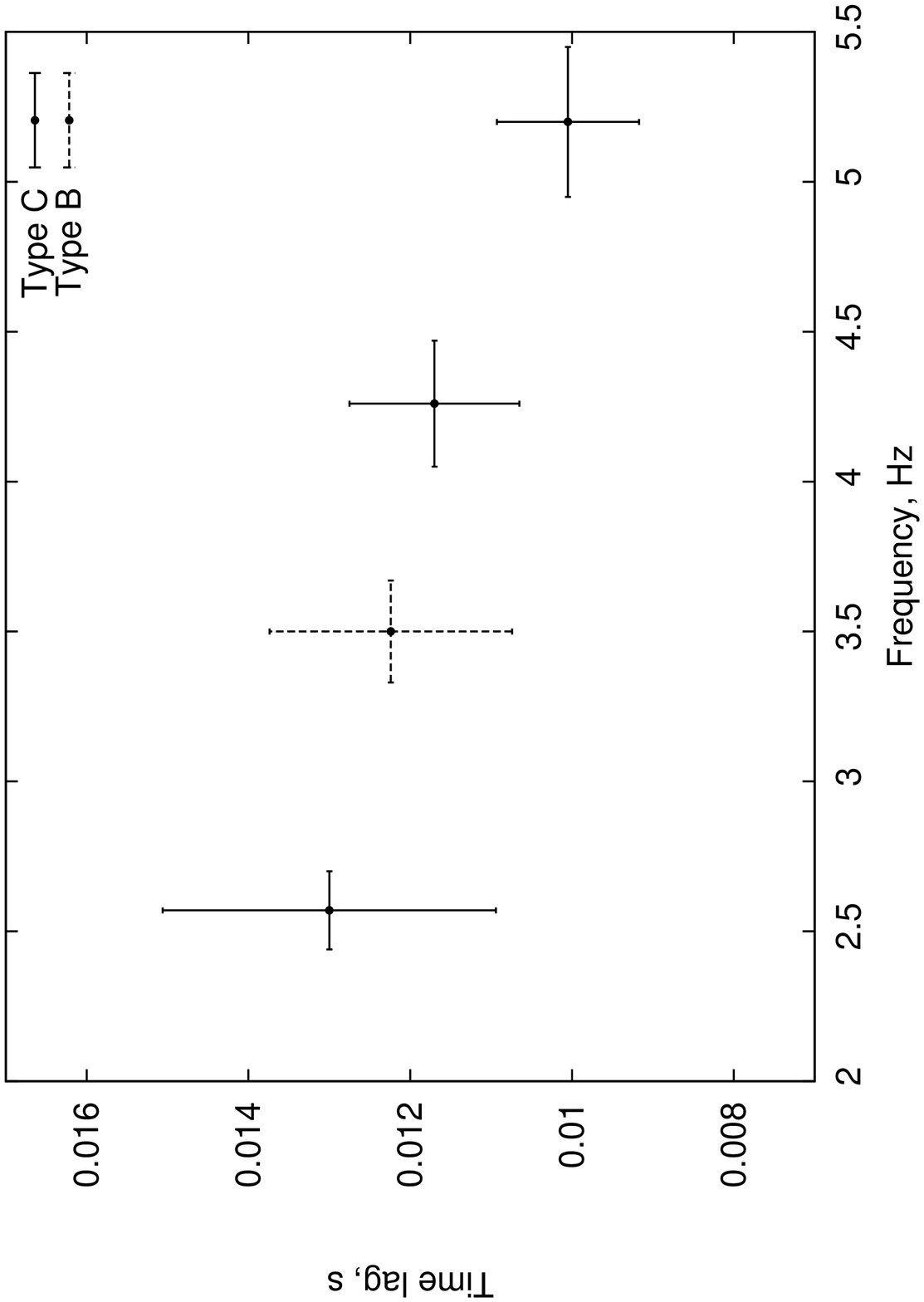}

\caption{ The QPO time lag versus the QPO frequency. The time lag is decreasing as the QPO frequency increases.
This is consistent with the lags being introduced by Comptonization in a collapsing corona. }
\end{figure}

In the LHS hard
time lags are associated with the aperiodic variability (see Mu{\~n}oz-Darias et al. 2010 for
the detailed time lag analysis during this observation), while the time lags in QPO during the transition
have become more pronounced and stay positive throughout
almost the entire transition episode. Only for the Type-A QPO  observation (Fig. 4, right panels)
the time lags become negative.  This is opposite to what was observed  in
XTE J1859-226  (Casella et al. 2004) and XTE J1550-564 (Remillard et al. 1999)
where the lags at the Type C QPO frequencies are negative
for the most part of the transition.  We note that in these observations QPO lags
appear to be affected by the negative lags introduced by the broad band noise 
component. Therefore, it is not conclusive that the intrinsic lags in the QPO
are negative and our analysis of J1752 data show clearly that at least in this
source the Type C QPO time lags are positive. In Figure 5  we plot the maximum time lag
observed in the QPO range versus QPO frequency. The decreasing trend in the time lag 
is apparent as the transition progresses. These observations can be 
qualitatively understood in terms of Comptonization process. Naturally,
if QPO  is a quasi-regular process in the Compton Corona, then
the time lag is the delay the hard photon should experience with respect to
the softer "seed" component during the up-scattering in the corona. As the corona
collapses, as expected in a standard state transition scenario, the time 
delay should decrease (e.g. Nowak et al. 1999). In fact, observations of both QPO frequency increase
and time delay decrease during state transition in J1752 are signatures of the
collapsing corona.

The negative time lags in the 95360-01-01-02 observation can be related
to a possible jet/outflow expected during this stage (Fender et al. 2009).
As the Compton corona becomes compact, the jet/outlow provides means
for the hard photons to be downscattered in energy and delayed with respect
to the hard emission. Alternatively, the soft delay in QPO can be introduced
by the perturbation propagation through the jet. 
While no major radio event during IS in J1752 is
reported so far, the radio brightness of XTE J1859-226 and XTE J1550-564
during transitions analyzed by Casella et al. and Remillard et al (see Fender et al. 
and references therein for details on radio observations of the galactic BHCs)
is consistent with  negative lags being
closely related to a jet or outflow (Shaposhnikov 2011, in prep). Note the hard rms spectrum 
of the Type A QPO (right bottom panel in Fig. 4). Negative time lags indicate
that the intrinsic spectrum of the matter involving in the QPO oscillations may be
even harder.

Similarly to the analysis presented above for J1752, in Figure 6 we present the timing
analysis of two observations of J1659. Hard lags seen in the first observation performed on Sept. 28, 2010 
are similar to the J1752 observation 94331-01-06-00 (Panel B in Fig. 3). Time lags are positive 
and follow approximately a power law distribution with frequency. There is no apparent contribution from
the QPO at 1.6 Hz.  The total rms spectra  show variability
increasing from 20 \% at 3 keV to 25 \% at 7 keV and then leveling off at higher energies. 
Separate rms spectra for the QPO and noise show different shapes.
The QPO rms distribution increases towards 10 keV and then saturates, similar to
the total rms. Oppositely, the noise rms spectrum  shows a soft distribution at energies below 10 keV
and then breaks to get harder above this energy. 
The second J1659 observation shown in Figure 6 was taken on Oct. 10
and shows a Type C QPO at 6.3 Hz . Notably, the rms spectra 
are similar to that shown by the J1752 observation 95360-01-01-02 
showing Type A QPO. The time lag associated with the QPO is clearly 
negative. 
The source behavior in 
radio wavelengths  reported by van der Horst et al. (2010) agrees with
in the jet-lag connection scenario proposed above (however, one should keep in mind
the radio quenching on Oct. 8).

\begin{figure}[t]
\centering
\hspace{-0.1in}\psbox[xsize=4.2cm,rotate=r]{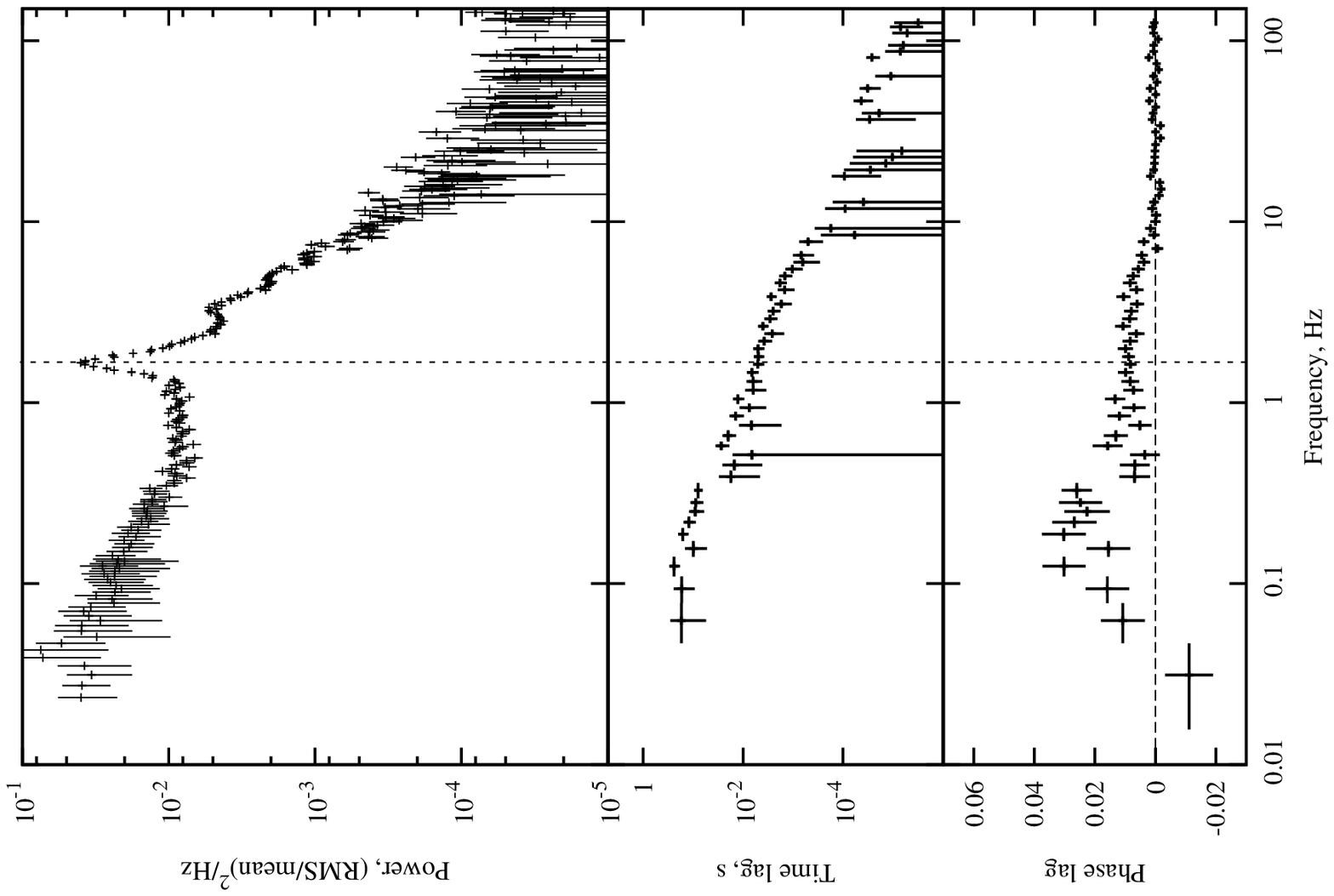}
\hspace{-0.1in}\psbox[xsize=4.2cm,rotate=r]{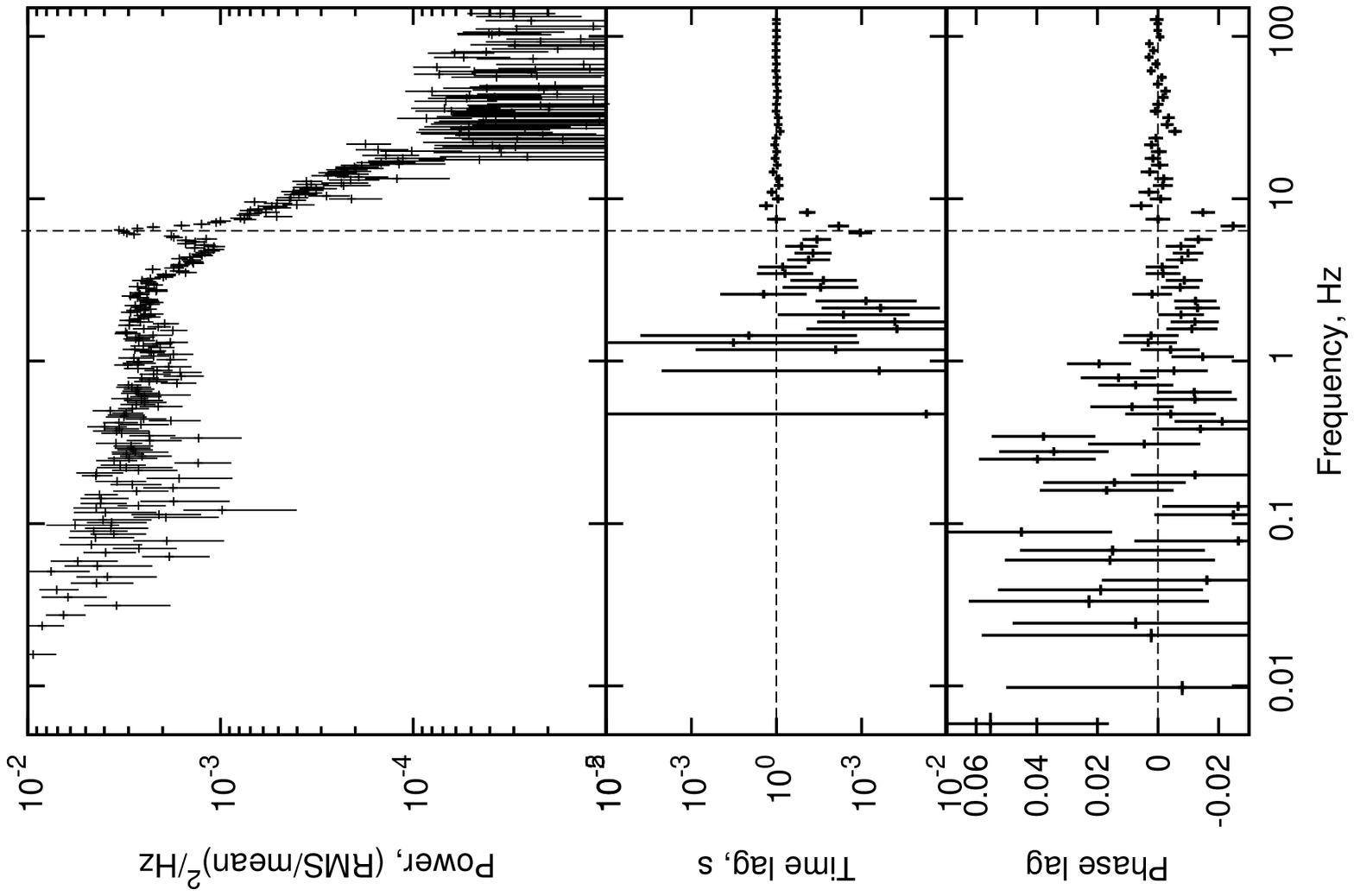}

\psbox[xsize=4.0cm,rotate=r]{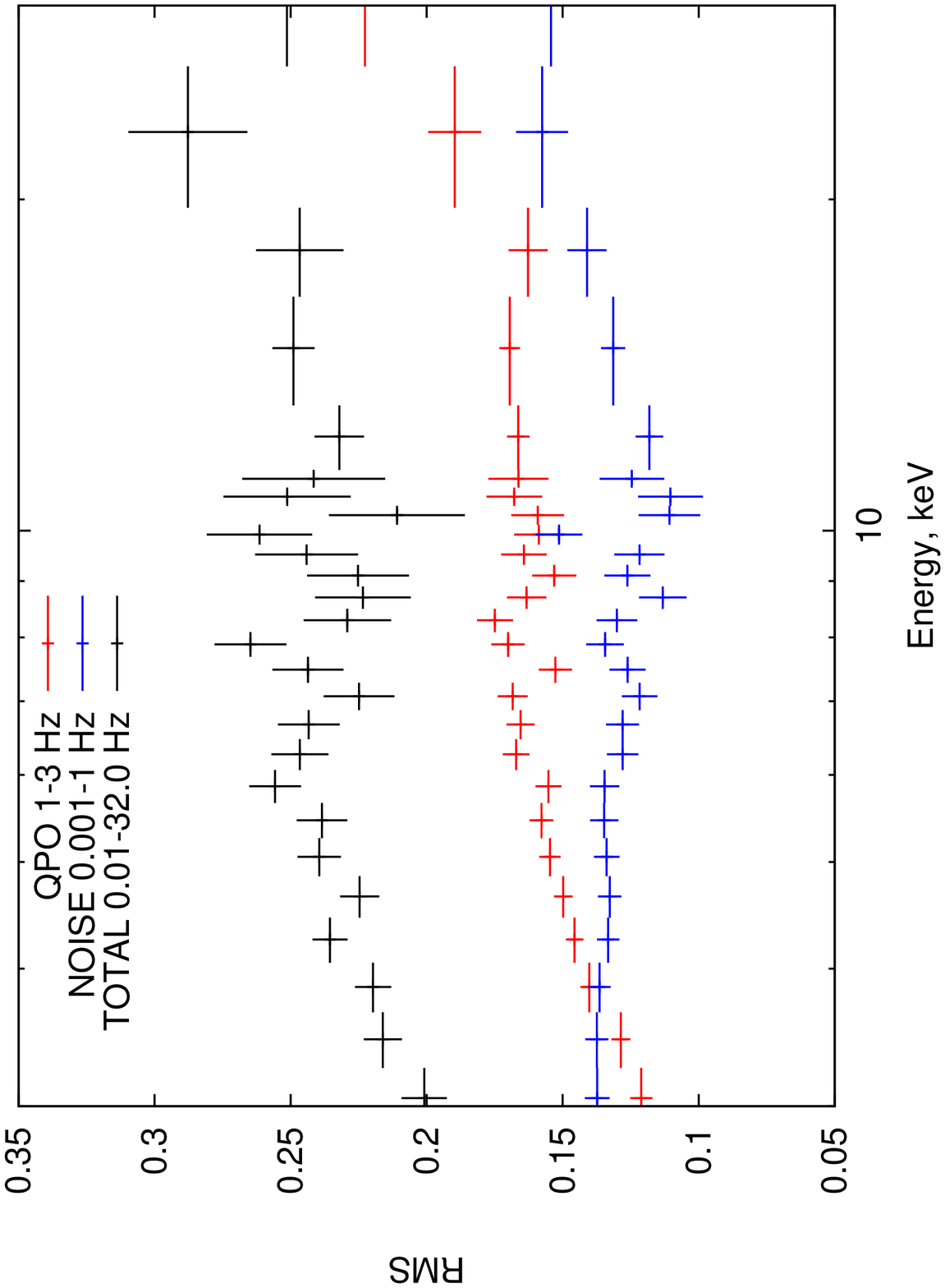}
\psbox[xsize=4.0cm,rotate=r]{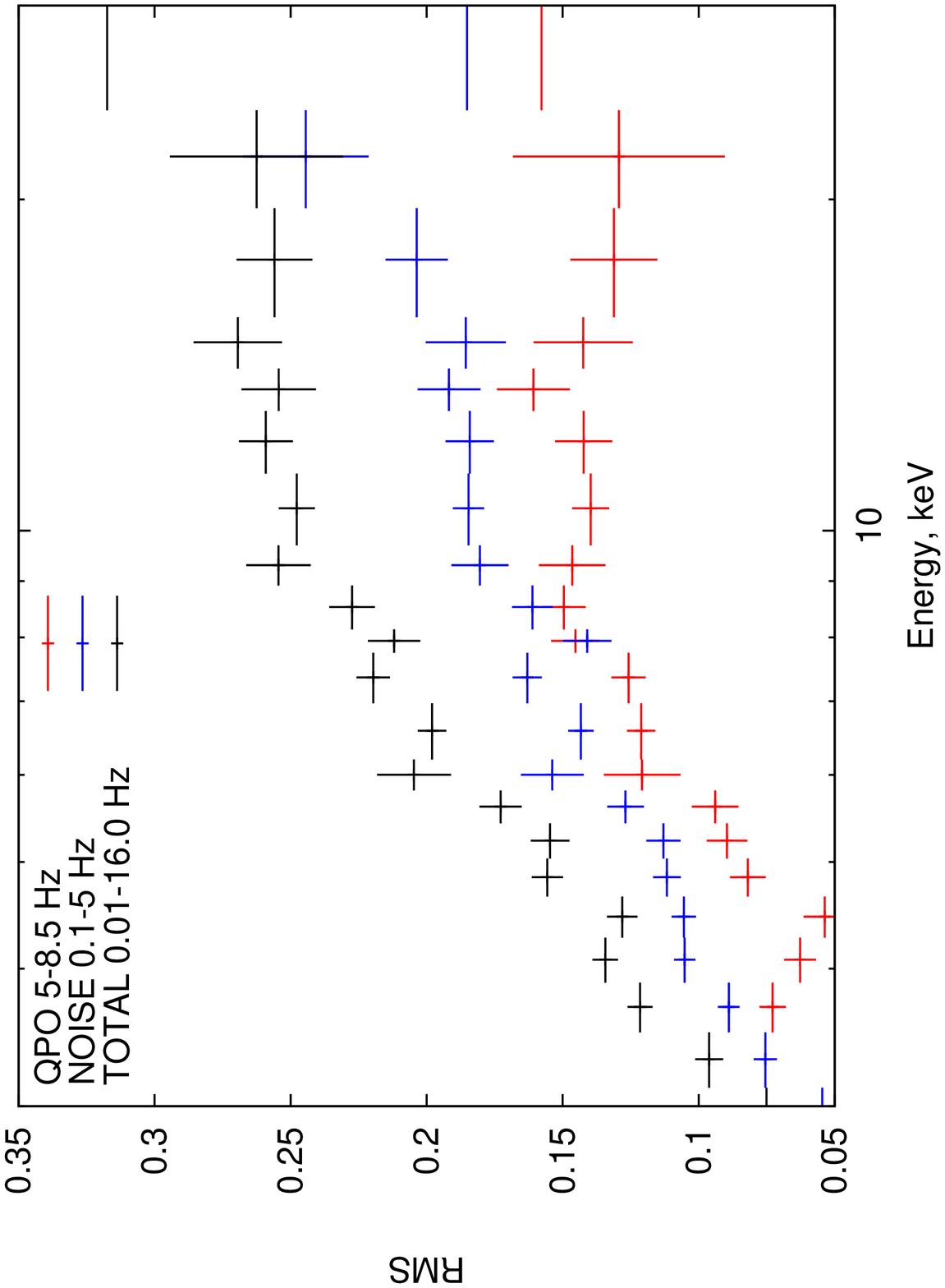}
\caption{ Same as Figure 3 for J1659 observations 95358-01-02-00 (left panels, the first {\it RXTE} observation of the source, start time 2010-09-28 00:43 UT) and 95108-01-22-00 (right panels, 2010-10-10 03:36 UT) both taken when the source was in the hard IS state.}
\end{figure}

\section{BH Masses in XTE J1752--223 and MAXI J1659--152}
\label{masses}

We use the mass determination method based on the scaling between spectral index and QPO frequency.
Combined with the information on the spectrum normalization, the method also can constrain the
distance to the source, albeit with some systematic uncertainty due to the unknown geometrical
factor (see ST09). BH mass for J1752 of 9.5$\pm$1.0 $M_\odot$ obtained by 
scaling  was reported by Shaposhnikov et al. 2010.  Here we apply scaling to J1659. 
The QPO-index correlation observed during the initial hard-to-soft transition has saturated at the 
index value of 2.3. We were unable to find a scalable reference pattern as all available rising transitions 
either 
have index saturation values exceeding 2.4 or level off below 2.2. We therefore use the decay data because
 decay transitions have similar index saturation value about 2.0$\sim$2.1 and usually are scalable. 
 By scaling the correlations observed in J1659 during the outburst 
decay to the 2003 decay transition data from GX 339-4 we obtain scaling coefficients 
$s_\nu = \nu_{GX 339-4}/\nu_{J1659} = 1.62 \pm 0.03$ 
and $s_N = N_{GX 339-4}/N_{J1659} = 1.08 \pm 0.03$. Using the BH mass and distance of GX 339-4 of
$M_{GX 339-4}/M_\odot=12.3\pm1.4$ and $d_{GX 339-4} = 5.8 \pm 0.7$ kpc,
obtained by ST09,  we infer the parameters for J1659 as $M_{J1659}/M_\odot=20\pm3$ and $d_{J1659} = 7.6 \pm 1.1$ kpc.
Here we assumed the geometrical factor of unity. Due to the fact that dips were observed in J1659 by {\it Swift} 
(Kuulkers et al. 2010) J1659 is probably has a high inclination. Therefore,
the distance to J1659 given above should be considered an upper limit.
The BH mass in J1659 presented above exclude any possibility
other than the source being an astrophysical BH. 

In Figure 8 we show the Index-QPO frequency correlation for the hard-to-soft transition in J1659.
The correlation is marked by a clear index saturation for high frequency values. This effect is proposed
as a signature of a BH, based on observation of a large set of galactic BH sources as well as
theoretical arguments (ST09). The index saturation seen in J1659 is in fact one of the most
pronounced among galactic BH candidates. Kalamkar et al. (2010b) classified J1659 as a BH candidate
using  an empirical argument based on the similarity with other BH sources. The index-QPO saturation
in J1659 is physically motivated evidence that J1659 is, in fact, a BH. The BH mass of 20 solar masses 
along with possible period of $\sim$2.5 hours makes J1659 the shortest period X-ray binary harboring
the heaviest BH. If confirmed, these results may have strong implications to the evolutionary
scenarios  of binary stars (see e.g. de Mink, for the discussion of the role of rotational mixing in producing 
such a heavy and compact binaries).



\section{Conclusions}
\label{summary}

\begin{figure}[t]
\centering
\psbox[xsize=8.5cm,rotate=r]{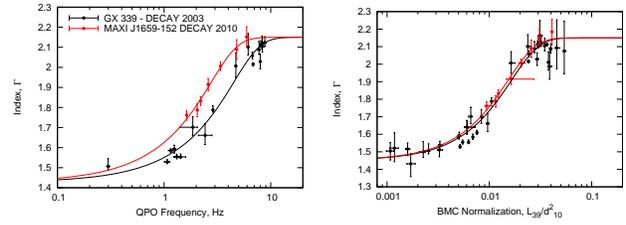}

\caption{ BH mass determination in J1659 with scaling method. We use the  
GX 339-4 data for the decay stage of the 2003 outburst (see ST09). 
On the left: Index versus QPO frequency. On the right: index versus the BMC model normalization.}
\end{figure}

\begin{figure}[t]
\centering
\psbox[xsize=8.5cm,rotate=r]{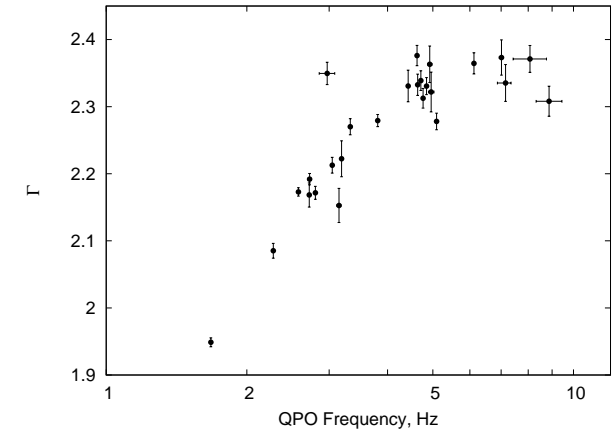}

\caption{ The Index - QPO frequency correlations during the hard IS of J1659.}
\end{figure}

We present detailed  comprehensive analysis of {\it RXTE} data collected during the discovery outbursts 
from new galactic BH binaries XTE J1752--223 and MAXI J1659--152. We show that the hard time lags observed
at the QPO frequency during the hard IS are related to delay during upscattering in Compton Corona, while negative lags 
observed later on, during the soft IS, may be explained by downscattering in a jet or outflow.

We also reportth BH mass determination in the two sources. While the BH in J1752 is close to the most other
BHs with a mass of about 9.5 $M_\odot$,  J1659 has BH mass of 20 $M_\odot$, making it the heaviest stellar 
mass BH in the Galaxy.

NS acknowledges support for this work  by NASA grant NNX09AF02G.

\section*{References}

\re
Arnaud, K.A., 1996,ASP Conf. Series volume 101, 17

\re 
Belloni, T.  2005,  AIP Conference, 797, 197 (astro-ph/0504185)

\re
Casella, P., Belloni, T., Homan, J., \& Stella, L.\ 2004, A\&A, 426, 587

\re
de Mink, S.~E., Cantiello, M., Langer, N., 
\& Pols, O.~R.\ 2010, AIP Conference Series, 1314, 291

\re
Fender, R.~P., Homan, J., \& Belloni, T.~M.\ 2009, MNRAS, 396, 1370 

\re 
 Jahoda, K., et al. 2006, ApJS, 163, 401 

\re
Kalamkar, M., et al. 2010a, ATel 2881

\re
Kalamkar, M., et al. 2010b, astro-ph/1012.4330

\re 
Kuulkers, E., et al. 2010, Atel 2912

\re 
Mangano, V., et al. 2010 GCN \#11296

\re
Markwardt C.B., et al. 2009, Atel. 2258

\re 
Mu{\~n}oz-Darias, T., et al. 2010, MNRAS, 404, L94

\re
Negoro, H. et al. 2010, Atel. 2873

\re 
Nowak, M.~A., et al. 1999, ApJ, 510, 874

\re
Remillard, R.~A., et al. 1999, ApJ, 517, L127

\re 
Remillard, R.~A.,  McClintock, J. E. 2006, ARA\&A,44,49

\re
Revnivtsev, M., Gilfanov, M., \& Churazov, E.\ 1999, A\&A, 347, L23

\re 
Shaposhnikov, N., \& Titarchuk, L.\  2009, ApJ, 699, 453

\re 
Titarchuk, L., \& Shaposhnikov, N. 2010, ApJ, 724, 2, 1147

\re
van der Horst, A.J., et al.  2010, Atel 2918

\label{last}

\end{document}